\documentclass[letterpaper]{article} % DO NOT CHANGE THIS
\usepackage{aaai25}  % DO NOT CHANGE THIS
\usepackage{times}  % DO NOT CHANGE THIS
\usepackage{helvet}  % DO NOT CHANGE THIS
\usepackage{courier}  % DO NOT CHANGE THIS
\usepackage[hyphens]{url}  % DO NOT CHANGE THIS
\usepackage{graphicx} % DO NOT CHANGE THIS
\usepackage{natbib}  % DO NOT CHANGE THIS AND DO NOT ADD ANY OPTIONS TO IT
\usepackage{caption} % DO NOT CHANGE THIS AND DO NOT ADD ANY OPTIONS TO IT
\usepackage{algorithm}
\usepackage{algorithmic}

\usepackage{newfloat}
\usepackage{listings}
\usepackage{subfigure}
\usepackage{booktabs}
\usepackage{amsfonts,amssymb}
\usepackage{amsmath}
\usepackage{subfigure}
\usepackage{tipa}
\usepackage{newfloat}
\usepackage{listings}
\DeclareCaptionStyle{ruled}{labelfont=normalfont,labelsep=colon,strut=off} % DO NOT CHANGE THIS
\floatstyle{ruled}
\newfloat{listing}{tb}{lst}{}
\floatname{listing}{Listing}
\title{Meta-Learning Empowered Meta-Face: Personalized Speaking Style Adaptation
for Audio-Driven 3D Talking Face Animation}
\author{
   Xukun Zhou\textsuperscript{\rm 1}, Fengxin Li\textsuperscript{\rm 1}, Ziqiao Peng\textsuperscript{\rm 1}, Kejian Wu\textsuperscript{\rm 2} ,Jun He\textsuperscript{\rm 1}, Biao Qin\textsuperscript{\rm 1}, Zhaoxin Fan\textsuperscript{\rm 3,4}, Hongyan Liu\textsuperscript{\rm 5}
}
\affiliations{
    \textsuperscript{\rm 1}Renmin University of China\\
    \textsuperscript{\rm 2}XREAL,
    \textsuperscript{\rm 3}Beijing Advanced Innovation Center for Future Blockchain and Privacy Computing, Institute of Artificial Intelligence, Beihang University, Beijing, China 
    \textsuperscript{\rm 4}
Beijing Academy of Blockchain and Edge Computing, China
    \textsuperscript{\rm 5} Tsinghua University
}

\usepackage{bibentry}
\begin{document}

\maketitle

\section{Abstract}
%Audio driven 3d face animation has become an important part in reallife human generating  and has been widely used in live streaming, 3D animation production and game generation. However, previous works mainly focus on generating human faces under known personal talking style, ignoring the model's ability to adapt to new person. To enhance the model's ability for  quick adaptation to unseen persons, we propose MetaFaces, which involves a meta-face as a good initialization for all style of talking faces. To make the adaption more flexible, we add a neural process to build the relationship between existing samples and unseen samples. To alleviate the memory cost, a low rank matrix is also used for adaption to each person. Experiments shows that our method achieves state-of-the-art results compared with all existing methods.
Audio-driven 3D face animation is increasingly vital in live streaming and augmented reality applications. While remarkable progress has been observed, most existing approaches are designed for specific individuals with predefined speaking styles, thus neglecting the adaptability to varied speaking styles. To address this limitation, this paper introduces MetaFace, a novel methodology meticulously crafted for speaking style adaptation. Grounded in the novel concept of meta-learning, MetaFace is composed of several key components: the Robust Meta Initialization Stage (RMIS) for fundamental speaking style adaptation, the Dynamic Relation Mining Neural Process (DRMN) for forging connections between observed and unobserved speaking styles, and the Low-rank Matrix Memory Reduction Approach to enhance the efficiency of model optimization as well as learning style details. Leveraging these novel designs, MetaFace not only significantly outperforms robust existing baselines but also establishes a new state-of-the-art, as substantiated by our experimental results.

% Audio-driven 3D face animation has become a pivotal component in realistic human modeling, finding extensive applications in live streaming, 3D animation production, and game development. Despite the progress made in this field, most existing approaches focus on generating facial animations for specific individuals with known speaking styles, often overlooking the model's adaptability to new, unseen faces. To address this limitation, we introduce MetaFaces, a novel framework designed to enhance the model's capability for rapid adaptation to previously unseen persons. 

% Our approach centers around the concept of a meta-face, which serves as a robust initialization point capable of accommodating a wide range of speaking styles. Additionally, to increase the flexibility of adaptation, we integrate a neural process that dynamically establishes relationships between existing samples and new, unseen data. To further optimize the model's efficiency and reduce memory overhead, we employ a low-rank matrix approach for individual-specific adaptation. 

% Extensive experiments demonstrate that our method consistently outperforms existing techniques, achieving state-of-the-art results across multiple benchmarks. This advancement not only broadens the applicability of audio-driven 3D face animation but also sets a new standard for future research in this domain.

\section{Introduction}

Audio-driven 3D talking face animation has become increasingly prevalent in various sectors, including gaming \cite{lin2021meingame}, live streaming \cite{hu2021virtual}, and animation production \cite{richard2021meshtalk}. Leveraging advanced technologies such as 3D parametric models \cite{peng2023emotalk}, Neural Radiance Fields \cite{peng2024synctalk}, and Gaussian splatting \cite{cho2024gaussiantalker}, these methods have achieved significant success in achieving accurate lip synchronization and facial emotions. Nonetheless, the intricate relationship between facial expressions and accompanying audio still needs to be explored. In other words, the problem of speaking style adaptation still needs to be addressed.

% \begin{figure}[htbp]
% 	\centering
% 	\begin{minipage}{1\linewidth}
% 		\centering
% 		\includegraphics[width=1\linewidth]{AnonymousSubmission/picture/normal.pdf}
% 		\caption{Normal training process of audio driven 3d face animation.}
% 		\label{chutian1}%文中引用该图片代号
% 	\end{minipage}
% 	%\qquad
% 	%让图片换行，
	
% 	\begin{minipage}{1\linewidth}
% 		\centering
% 		\includegraphics[width=1\linewidth]{AnonymousSubmission/picture/meta.pdf}
% 		\caption{Training process of MetaFace.}
% 		\label{chutian2}%文中引用该图片代号
% 	\end{minipage}
% \end{figure}
\begin{figure}[ht]
	\centering
		\subfigure[Speaking style adaptation mode of existing methods.]{
  \includegraphics[width=0.45\textwidth]{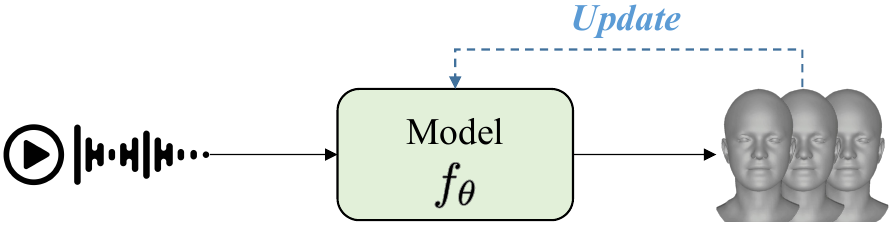}
  \label{fig.noraml}
  } \\
    	\subfigure[Speaking style adaptation mode of MetaFace.]{
     \includegraphics[width=.45\textwidth]{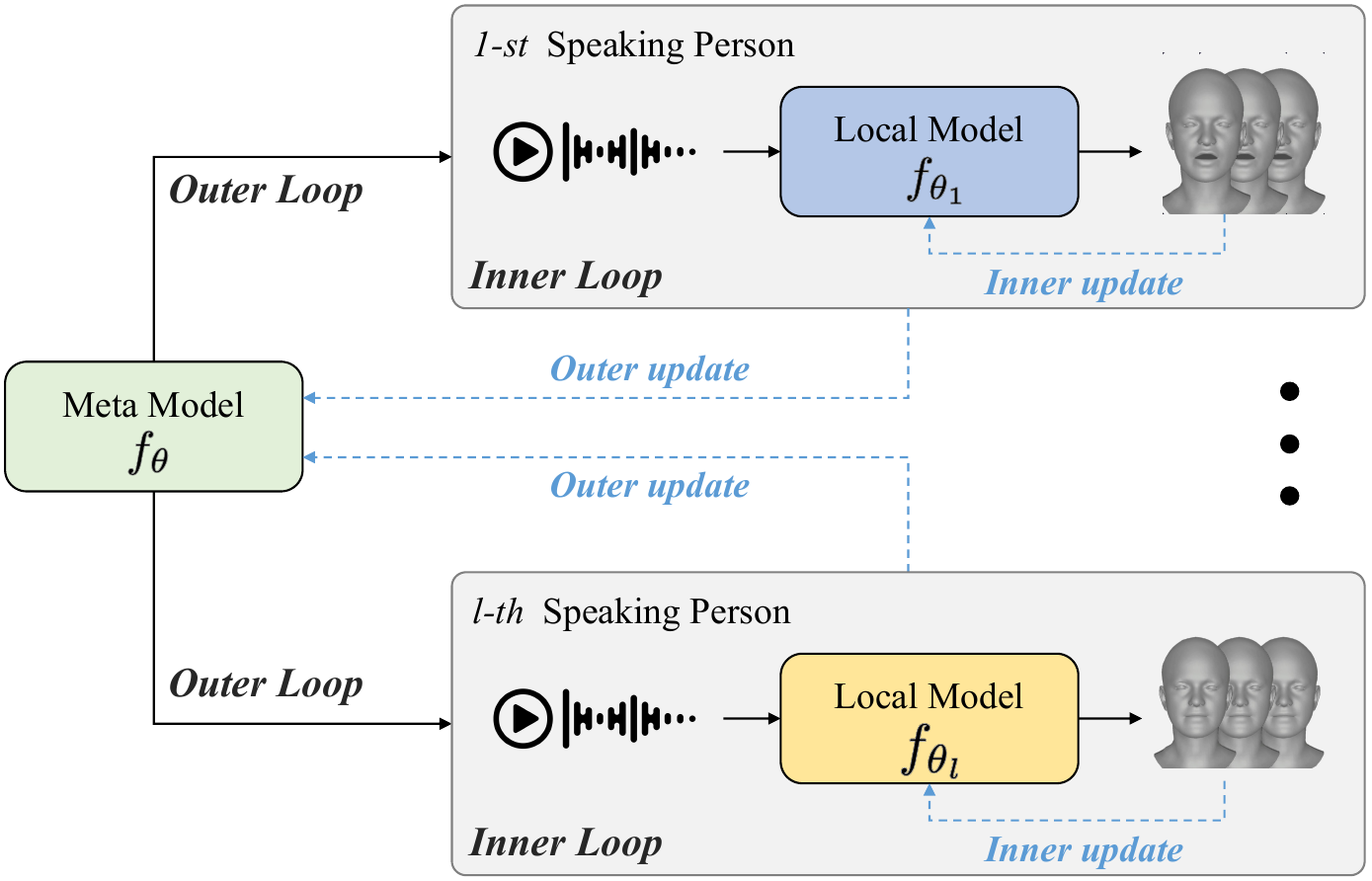}
     \label{fig.meta}
     }
     \vspace{-0.1in}
	\caption{Difference between speaking style adaptation mode of existing methods and MetaFace.}
 \vspace{-0.25in}

\end{figure}

Several methods have been proposed to address this issue in recent years \citep{cudeiro2019capture,fan2022faceformer,peng2023emotalk,peng2023selftalk}. For instance, SelfTalk \cite{peng2023selftalk} introduces a self-training pipeline that leverages text and 3D lip meshes for speaking style ad aptation. Meanwhile, FaceFormer \cite{fan2022faceformer} focuses on training different individuals with predefined speaking styles. Although these approaches mark some progress, they still face significant challenges: \textbf{1) Firstly}, most methods require a substantial amount of data \cite{fan2022faceformer,song2024talkingstyle} for effective speaking style distillation. \textbf{2) Secondly}, the prevalent use of cross-training technologies \citep{peng2023emotalk, style_cross} for speaking style adaptation necessitates paired sentences, reducing flexibility in application. Therefore, a pertinent question emerges: \emph{\textbf{Is it possible to devise a flexible and straightforward method for speaking style adaptation in audio-driven 3D face animation that utilizes minimal data?}}

%, which construct same human face motions with different condition such as emotions and person id,necessitate paired sentences, where different individuals speak the exact same phrase.

%Previous work\citep{cudeiro2019capture,fan2022faceformer,peng2023emotalk,peng2023selftalk} has focused heavily on distilling individual talking styles or emotional nuances into compact, low-dimensional representations and fine-tuning specific model layers to improve feature representation. These methods, however, face two major challenges. First, training a robust human representation embedding or a well-defined personal style typically requires a large volume of data\cite{fan2022faceformer,song2024talkingstyle}, which can be challenging to acquire. Second, the most commonly employed cross-training techniques\citep{peng2023emotalk, style_cross}, which construct same human face motions with different condition such as emotions and person id,necessitate paired sentences, where different individuals speak the exact same phrase.

To answer this question, we introduce MetaFace, a novel framework designed explicitly for Personalized Speaking Style Adaptation in Audio-Driven 3D Talking Face Animation. The core idea of MetaFace, illustrated in Fig. \ref{fig.meta}, is grounded in the principle that Audio-Driven 3D talking face animation can significantly benefit from the foundational concepts of meta-learning. Specifically, to facilitate rapid adaptation with minimal data, MetaFace consists of a Robust Meta Initialization Stage (RMIS) and a Low-rank Matrix Memory Reduction Approach for speaking style adaptation. In MetaFace, the RMIS is utilized to initialize the network weights from a pre-trained 3D talking face animation model, infusing the network with basic style cues of the target individual. Subsequently, the latter employs the renowned LoRA \cite{hu2021lora} fine-tuning strategy to equip the model with personalized details quickly. Furthermore, acknowledging the challenges that few-shot training often faces in handling training and test data from different domains separately, MetaFace introduces a Dynamic Relation Mining Neural Process (DRMP) to enable the model to establish additional connections between training and testing samples, thereby further enhancing the performance of speaking style adaptation.

Extensive experiments are conducted using the widely recognized VOCASet \citep{cudeiro2019capture} and BIWI \cite{BIWI} datasets. The experimental outcomes demonstrate that MetaFace outperforms existing state-of-the-art methods, showcasing its superior performance in the field.

Our contributions can be summarized as follows:
\begin{itemize}
    \item We introduce MetaFace, a novel framework designed explicitly for Personalized Speaking Style Adaptation in Audio-Driven 3D Talking Face Animation, leveraging the foundational principles of meta-learning.
    \item We develop key components within MetaFace, including the Robust Meta Initialization Stage, the Dynamic Relation Mining Neural Process, and the Low-rank Matrix Memory Reduction Approach, collectively enhancing the framework's performance to achieve state-of-the-art results.
    \item We rigorously evaluate our method against existing techniques, demonstrating its superior effectiveness through extensive experiments conducted on several widely recognized datasets.
\end{itemize}

\section{Related Work}
This paper primarily focuses on audio-driven 3D talking face animation, explicitly emphasizing the problem of personalized speaking style adaptation. To this end, this section first reviews related works concerning Lip Synchronization in 3D Face Animation. Subsequently, it discusses contributions in Style Speaking Adaptation in 3D Face Animation. Finally, we explore the role of Meta-learning and Parameter Adaptation, which serve as foundational technologies in the MetaFace framework.

\subsection{Lip Synchronization in 3D Face Animation}
For most 3D face animation techniques, the primary objective remains the achievement of precise lip synchronization. Early methods \citep{ezzat1998miketalk, ezzat2000visual} primarily rely on audio produced by Text-to-Speech technologies \citep{black1998festival} and real-life facial motion videos animated using visemes \citep{fisher1968confusions} to extract features synchronized with audio. Although these methods are relatively straightforward, they depend heavily on meticulously organized datasets, posing significant challenges in creating animations tailored to individual users. To overcome these limitations, deep learning-based approaches \citep{karras2017audio, fu2024mimic, peng2023emotalk, danvevcek2023emotional, peng2023selftalk} have been introduced. \citet{karras2017audio} pioneers using neural networks to generate mesh movements directly from audio features. Subsequently, \citet{richard2021meshtalk} introduces a method that separates audio-correlated and uncorrelated information, yielding more realistic animation performance. Rather than investigating audio features, \citet{fan2022faceformer} focuses on the prediction mode, enhancing the efficiency of transformer models in facial motion prediction by employing an auto-regressive model. Building on this foundation, \citet{xing2023codetalker} further develops a code query-based approach to speech-driven 3D facial animation, improving realism by minimizing mapping uncertainties and outperforming existing methodologies. Recently, inspired by lip reading techniques, \citet{peng2023selftalk} further enhances the accuracy of lip movements by integrating audio, textual content, and lip shapes, achieving state-of-the-art results.

Although the methods mentioned above have achieved remarkable progress, they primarily focus on improving lip synchronization, while the equally important question of personalized speaking style adaptation has been largely overlooked. In our work, we introduce MetaFace to address this issue in both an effective and efficient manner.

\subsection{Speaking Style  Adaptation in 3D Face Animation}

Recent advancements in 3D Face Animation have increasingly focused on adapting speaking styles. Studies primarily explore two approaches: utilizing external personality labels \citep{cudeiro2019capture,tian2019audio2face,fan2022faceformer,peng2023emotalk}, and decoupling facial features from audio \citep{style_cross,peng2023emotalk}. The first approach employs personality labels to foster the learning of personalized embeddings, exemplified by \citet{cudeiro2019capture}, who created a personalized dataset and a model to integrate individual styles. This foundation has spurred methods that incorporate emotional aspects \citep{danvevcek2023emotional} or specific speaking styles \citep{thambiraja2023imitator,fan2022faceformer,song2024talkingstyle}. The second approach focuses on disentangling speaking style from audio features, with innovations like cross-sample training to separate emotional and content features \citep{peng2023emotalk,style_cross,fu2024mimic}. \citet{zhang20213d} and \citet{song2024talkingstyle} further enhance this by utilizing identity and facial motions as additional training conditions, improving model performance even with minimal data \citep{thambiraja2023imitator}.

Although these approaches mark significant progress, they still face substantial challenges. Firstly, most methods necessitate a considerable amount of data for effective speaking style distillation. Additionally, the prevalent use of cross-training technologies for speaking style adaptation requires paired sentences, reducing application flexibility. MetaFace addresses these issues by leveraging meta-learning and low-rank fine-tuning principles.

\subsection{Meta-learning and Parameter Adaptation}
Meta-learning is a branch of machine learning methods that enables deep models to adapt to novel categories with minimal training data \citep{hospedales2021meta}. The concept of meta-learning was first introduced by \citet{wirth2008learning}, which explored adapting the learning rate for novel category adaptation. Subsequently, MAML \cite{finn2017model} proposed a model-agnostic meta-learning approach for fast adaptation by utilizing the concept of learning to learn. Following this, numerous works have been proposed to enhance this line of research from diverse perspectives \cite{huisman2021survey}. In parallel, advancements in model pretraining have significantly influenced meta-learning. Initially introduced by \citet{devlin2018bert}, pretraining models on large datasets and fine-tuning them on smaller tasks has been influential across various domains \citep{wang2022pre,du2022survey}. However, with the advent of large language models like GPT-3 and GPT-4 \citep{gpt3,gpt4}, fine-tuning entire models for small tasks has become computationally prohibitive. To address this limitation, \citet{hu2021lora} proposes a solution by decomposing model parameters into two low-rank matrices, significantly reducing computational costs while maintaining performance. This approach, known as LoRA, has been widely adopted across various applications \citep{wang2024prolificdreamer,dettmers2024qlora,zhang2023adding}.

Although numerous meta-learning methods and parameter adaptation techniques have been proposed and widely applied across various domains, their potential in efficient 3D face animation still needs to be explored. In this paper, we adopt the concept of meta-learning and low-rank fine-tuning for speaking style adaptation in 3D talking face animation.

\section{Method}
\subsection{Overview}
%In this section, we would describe our model, MetaFace, in detail. Our main idea is that the audio-face animation methods have a common relation-ship and could be a basic model for personalized face.  Our model could be divided into two part:  Face Regression module, Neural process based Feature  and a low-rank matrix finetuning module, shown as Fig.\ref{fig.structure} During training, we use model agnostic  learning method \citep{finn2017model} to get a good global initiation and finetune with several samples on specific task samples. 

% In this section, we will provide a detailed description of our model, MetaFace. Our central idea is that audio-face animation methods share a common relationship, which can serve as a foundational model for personalized face animation. Our model is divided into three main parts: the Robust Meta Initialization Stage and the Dynamic Relation Mining Neural Process, along with Low-rank Matrix Memory Reduction Approach, as shown in Fig. \ref{fig.structure}. During training, we utilize the Model-Agnostic Meta-Learning (MAML) method \citep{finn2017model} to achieve a robust global initialization named "meta face", which is then fine-tuned using several samples specific to the task at hand.

This work explores the critical issue of personalized speaking style adaptation. Contrary to existing supervised learning frameworks such as those proposed by \cite{fan2022faceformer}, \cite{peng2023emotalk}, and \cite{song2024talkingstyle}, which delineate the mapping relationship between audio and facial animation for all speakers within a dataset, our study advocates for a "meta-face" methodology, tailored explicitly for superior adaptation to novel individuals.

Specifically, we first examine an audio-driven 3D facial animation dataset assembled from a diverse group of individuals, denoted as $\mathcal{P}=\{1,2,\cdots,p,\cdots,|\mathcal{P}|\}$, where $p$ identifies the $p$-th participant. For each individual $p$, the dataset encapsulates observed audio-face pairings, represented as $\mathcal{D}_p=\{(\mathbf{a}_{p, m}, \mathbf{v}_{p, m}), m \in \{1,2,\cdots,|\mathcal{D}_p|\}\}$. Here, $\mathbf{a}_{p, m} \in \mathbb{R}^{T_a}$ specifies the audio sequence of the individual, spanning a duration of $T_a$. Concurrently, $\mathbf{v}_{p, m} \in \mathbb{R}^{T_v \times L \times 3}$ describes the corresponding facial motion sequence, aligned to a reference template face $T \in \mathbb{R}^{L \times 3}$. The parameters $T_v$, $L$, and $3$ denote the length of the facial sequence, the number of facial landmarks, and the spatial coordinates $(x, y, z)$, respectively. The collective dataset is represented as $\mathcal{D}=\{\mathcal{D}_p, p \in \mathcal{P}\}$.

The final objective is to construct a model, denoted by $f_\theta$, by leveraging the samples from the dataset $\mathcal{D}$. This model is subsequently refined through fine-tuning with the personalized dataset $P_l$. This adaptation process enhances the model's efficacy $f_{\theta_l}$ on the dataset specific to individual $l$. More formally, for a given task $i$, the query and support set are defined as $\omega_i = \{(\mathbf{a}_{k}, \mathbf{v}_{k}), k \in \{1,2,\cdots,K_i\}\}$, with $K_i$ representing the count of samples within $\omega_i$.

\begin{equation}
\theta^* = \arg\min_{\theta} \sum_{(\mathbf{a}, \mathbf{v}) \in \mathcal{D}} \mathcal{L}(f_\theta(\mathbf{a}), \mathbf{v})
\end{equation}

Upon obtaining the optimal parameters $\theta^*$, the model is further personalized as follows:

\begin{equation}
\small
\theta_l^* = \arg\min_{\theta} ( \lambda \sum_{(\mathbf{a}, \mathbf{v}) \in P_l} \mathcal{L}(f_{\theta}(\mathbf{a}), \mathbf{v}) + (1 - \lambda) \mathcal{L}(f_{\theta^*}(\mathbf{a}), \mathbf{v}) )
\end{equation}

, where $\lambda$ is a tuning parameter that balances the influence of the personalized data $P_l$ against the pre-trained model parameters.
\begin{figure*}[htbp]
\centering
\includegraphics[scale=0.3]{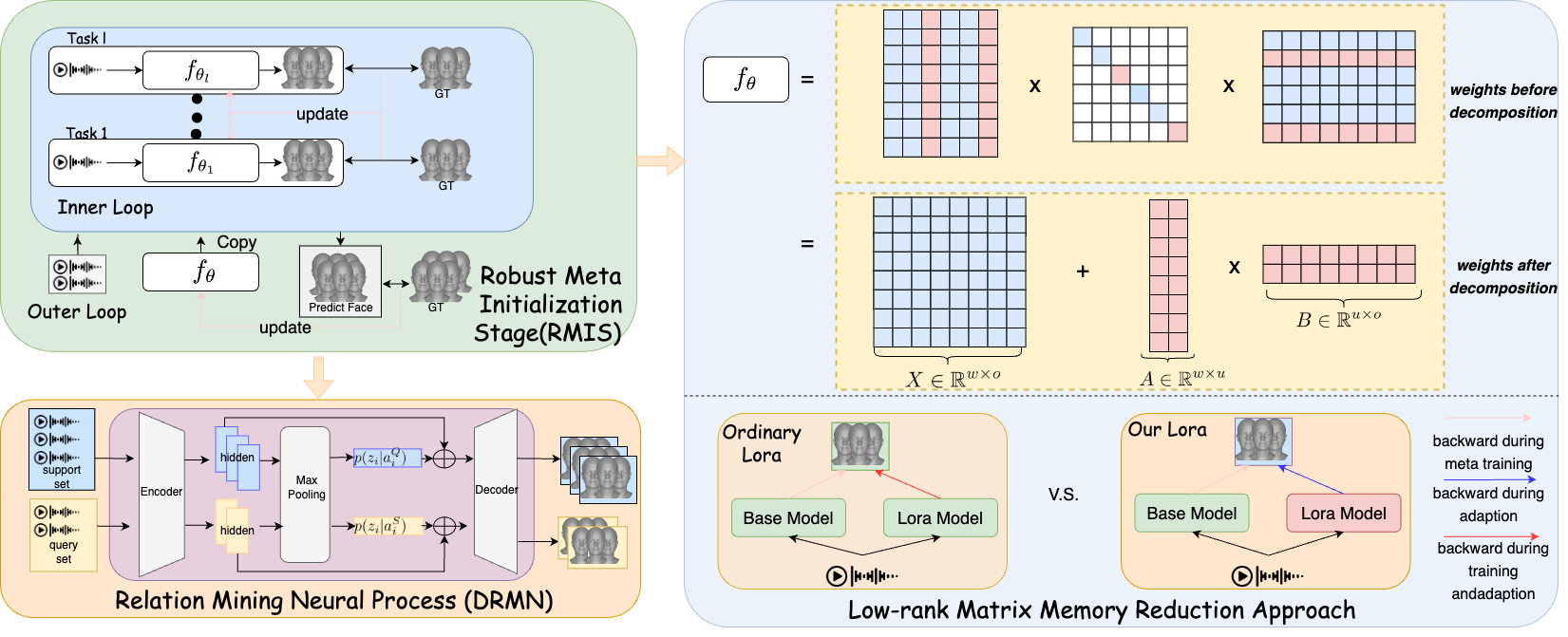}
\vspace{-0.1in}
\caption{Overall Framework of MetaFace.} 
\label{fig.structure}
\vspace{-0.2in}
\end{figure*}
Fig. \ref{fig.structure} demonstrates the operational framework of MetaFace. Starting with a pre-trained 3D talking face animation model alongside a minimal training dataset, MetaFace initially engages the Robust Meta Initialization Stage to expedite the model weight updates. This stage allows the model to rapidly assimilate basic cues from a new individual efficiently. Additionally, this phase incorporates a Dynamic Relation Mining Neural Process into the 3D face animation model. This enhancement facilitates the model's ability to discern the domain disparities between previously modelled individuals and the new subject and to unearth intrinsic relationships, thereby promoting model adaptability to the speaking style domain of the new individual. Subsequently, the primed deep model undergoes refinement through the Low-rank Matrix Memory Reduction Approach. This method enhances the LoRA \cite{hu2021lora} technique, further aiding the model in acquiring intricate details of novel speaking styles.

The following subsections delve into the intricacies of the Robust Meta Initialization Stage, the Dynamic Relation Mining Neural Process, and the Low-rank Matrix Memory Reduction Approach.

% The objective of audio-driven 3D facial animation is to generate the facial motion $\hat{\mathbf{v}}_{p, m}$ based on the input audio $\mathbf{a}_{p, m}$ and the individual's identity $p$, formulated by:
% \begin{align}
%     \hat{\mathbf{v}}_{p, m} = f_{\theta}(\mathbf{a}_{p, m}, p)
% \end{align}
% where $f_{\theta}$ represents the generation function with parameters $\theta$. 
% Task $i$ query set and support set 
% $\omega_i = \{(\mathbf{a}_{k}, \mathbf{v}_{k}), k \in {1,2,\cdots,K_i}\}$, $K_i$ denotes the amount of the samples in $\omega_i$

%Specifically, let $D=\{(a_s,v_s) | a_s \in A_s,v_s \in V_s\}$ where $v_t \in \mathbb{R}^{n\times 3}$. represent the audio and face motion data in the support dataset. Here, $T$ is the sequence length and $
%n$ is the number of vertices on the face. Additionally, we define the personalized dataset $P_l=\{(a_l,v_l|a_l \in A_l ,  v_l \in V_l\}$ where $A_l$ and $V_l$ are small personalized dataset of person $l$. 
% Our goal is to train a model 
% $f_\theta$
%   using samples from dataset 
% $D$, which can then be fine-tuned with the personalized dataset $P_l$. Through this adaptation, the model $f_{\theta_l}$
%   will perform better on the personalized dataset of person $l$.task $i$ query set $\omega_i = \{(\mathcal{D}_p , p \in \mathcal{S}_i \cup \mathcal{Q}_i\}$
% and support set 
% $\omega_i = \{(\mathbf{a}_{k}, \mathbf{v}_{k}), k \in {1,2,\cdots,K_i}\}$, $K_i$ denotes the amount of the samples in $\omega_i$

\subsection{Robust Meta Initialization Stage}
\label{meta_learning}

To equip a pretrained 3D talking face animation model with the capability to learn an individual's unique speaking style, we begin with the Robust Meta Initialization Stage. This stage updates the model's weights to produce an initial weight configuration. We define the audio input as $a$ and the face animation bias as $v$, while the speaking feature distribution for person $l$ is represented as $p(l)$.

Given that traditional training approaches, which compute the loss $\tau$ using the model $f_{\theta}$ over the entire batch, may not adequately address unseen speaking styles, we employ a strategy inspired by \citet{finn2017model}. This involves training a meta face model as a robust initialization for adapting to new samples.

For a new individual $l$, we construct a person-specific dataset $\mathcal{D}_l$ containing $K_l$ samples drawn from the distribution of speaking features of person $l$. Each sample consists of audio $a$ and face motion bias $v$. The model is initially updated using the gradients calculated from the performance of model $\theta$ on these $K_l$ samples, as indicated in Equation \ref{equ.update}:

\begin{equation}
    \hat{\theta}_p = \hat{\theta}_l - \alpha \nabla_{\hat{\theta}} \sum_i \tau(f_{\hat{\theta}_l}(a_l, l), v_l)
    \label{equ.update}
\end{equation}

The parameters $\hat{\theta}_p$, updated locally, are deemed optimal for the dataset $\mathcal{D}_p$. To verify the precision of these parameters, we resample another $K$ observations $\phi'$ from $p(l)$, which are used as a query set to compute the gradient through the loss $\tau(f_{\hat{\theta}}(a', l), V')$. Subsequently, the original model parameters $\theta$ are refined using the following update rule:

\begin{equation}
    \theta = \theta - \beta \nabla_{\theta} \sum_{l=1}^{K} \tau(f_{\hat{\theta}_l}(a_l, l), v_l)
    \label{equ.maml}
\end{equation}

,where $\beta$ is the learning rate. Detailed procedures of  meta-face algorithm are delineated in Algorithm \ref{alg3}.

\subsection{Low-rank Matrix Memory Reduction Approach}
Following the Robust Meta Initialization Stage, the subsequent phase involves the meticulous model fine-tuning. This crucial step enables the model to assimilate additional, intricate details specific to each individual. The fine-tuning process is pivotal as it refines the model's ability to capture and replicate the nuanced characteristics inherent in the speaking style of the subject, thereby enhancing the overall adaptability and effectiveness of the model. Given the impracticality of maintaining a fully adaptive model for each individual, we adopt a strategy that leverages a low-rank decomposition approach, as suggested by \citet{hu2021lora}, to reduce the memory footprint significantly. Specifically, for the linear layers' weight matrix $\mathbf{X} \in \mathbb{R}^{w \times o}$ within the model parameters $\theta$, we apply a low-rank decomposition. Traditionally, updates to the linear parameters are performed as $\mathbf{X} = \mathbf{X} + \Delta\mathbf{X}$. However, in our approach, these updates are replaced by $\mathbf{X} = \mathbf{X} + \Delta\mathbf{B}\mathbf{A}$, where $\mathbf{B} \in \mathbb{R}^{w \times u}$ and $\mathbf{A} \in \mathbb{R}^{u \times o}$ are the low-rank matrices and $u$ denotes the predefined dimensions, significantly less than $w$ or $o$.

\begin{equation}
    \mathbf{X}_{new} = \mathbf{X} + \Delta\mathbf{B}\mathbf{A}
\end{equation}

For the entire model's parameters $\theta$, we construct a low-rank product $\theta_m$ to update the personality network. To enhance the adaptive capacity of the model concerning individual speaking styles, we propose that the LoRA parameters $\theta_m$ should also be concurrently trained with $\theta$ during the meta-learning phase. This dual training process involves $\theta$ and $\theta_m$ when training the meta face. 

\begin{equation}
    \theta_{final} = \text{Train}(\theta, \theta_m)
\end{equation}

Notably, while traditional approaches typically initialize the LoRA parameters randomly during fine-tuning, we innovate by initializing $\theta_m$ using meta-learning techniques. This initialization contributes to memory reduction and enhances the model's ability to capture detailed nuances of the speaking style. During the fine-tuning phase on a specific individual, only the $\theta_m$ parameters are updated, thereby maintaining a low memory footprint while fine-tuning to adapt to the unique characteristics of the individual's speaking style.

% 假设  任务是 p_l  用 l 表征一个人  
% \subsection{Neural Process for Meta-Training}
% Conventional meta-learning methods regard each individual as an independent task. This mechanism neglects the face correlations among different persons, where the similar persons may have similar face motions. This correlation could provide beneficial for training the meta parameters. To mining the correlation, we propose a Neural Process for Meta-Training (NPMT).

%\subsection{Notations for Meta-Training}

\begin{algorithm} 
	\caption{Learning methods of 
Robust Meta Initialization Stage.} 
	\label{alg3} 
	\begin{algorithmic}
		\REQUIRE Dataset $D$ contains $P$ person's speaking audio $a$ and face motion bias $b$.
        \REQUIRE Learning rate $\alpha$ and $\beta$
		\STATE Random initialize $\theta$
		\WHILE{Not Done}
        \FOR{ Person $l$ in dataset $D$}
		\STATE Sample $K$ samples $\phi_l=(a_l,V_l)$ from speaking style distribution $p(l)$
        \STATE Evaluate model $\theta$'s performance  $\tau(f_\theta(a_l,l),V_l)$
        \STATE Compute adapter parameter 
        $\hat{\theta_l}=\theta-\alpha\bigtriangledown_\theta\tau(f_\theta(a_l,l),V_l)$
        \STATE Sample new observation $\phi_l'=(a_l',V_l')$ from distribution $p(l)$
        \STATE Compute update $\tau(f_{\hat{\theta_l}}(a_l',l),V_l')$
        \ENDFOR
        \STATE Update the model parameter $\theta = \theta - \beta\Delta_\theta \sum_{l=1}^{K}\tau(f_{\hat{\theta_l}}(a_l',l),V_l')$ 
        
		\ENDWHILE 
	\end{algorithmic} 
\end{algorithm}

\subsection{Dynamic Relation Mining Neural Process}

In the Robust Meta Initialization Stage, as detailed in Section \ref{meta_learning}, the updating rule considers each individual a distinct task. This approach inadvertently neglects the potential facial correlations among individuals, where similarly characterized persons might exhibit akin facial movements. Harnessing these correlations could significantly bolster the training of the meta-parameters. We have introduced the Dynamic Relation Mining Neural Process to tap into and exploit these correlations.

Specifically, for each task $\omega_i = \{\mathcal{D}_l , l \in \mathcal{S}_i \cup \mathcal{Q}_i\}$, we postulate that $\omega_i$ emanates from a stochastic process $h_i$, with each data point $(\mathbf{a}_{k}, \mathbf{v}_{k})$ epitomizing a sample from $h_i$. The conditional probability distribution $p(\mathbf{v}_{1:K_i}|\mathbf{a}_{1:K_i})$ can be delineated as follows:
\begin{align}
    p(\mathbf{v}_{1:K_i}|\mathbf{a}_{1:K_i}) = \int p(h_i) p(\mathbf{v}_{1:K_i}|\mathbf{a}_{1:K_i}, h_i) d h_i
\end{align}
where $\mathbf{a}_{1:K_i}$ and $\mathbf{v}_{1:K_i}$ represent all samples $(\mathbf{a}_{k}, \mathbf{v}_{k})$ from task $\omega_i$, respectively, and $K_i = |\omega_i|$ denotes the count of samples in $\omega_i$. The Neural Process approximates the stochastic process $h_i$ using a random vector $p(\mathbf{z}_i)$, reformulating the equation as:
\begin{align}
    p(\mathbf{v}_{1:K_i}|\mathbf{a}_{1:K_i}) = \int p(\mathbf{z}_i) \prod^{K_i}_{k=1}p(\mathbf{v}_{k}|\mathbf{a}_{k}, \mathbf{z}_i) d \mathbf{z}_i
\end{align}
Given the intractability of the posterior distribution, the Neural Process employs variational inference, introducing a variational posterior $q(\mathbf{z}_i|\omega_i)$. The Evidence Lower-BOund (ELBO) is thus derived as:

\begin{equation}
\scriptsize
    \log p(\mathbf{v}_{1:K_i}|\mathbf{a}_{1:K_i}) \geq \mathbb{E}_{q(\mathbf{z}_i|\omega_i)} \left[ \sum_{k=1}^{K_i} \log p(\mathbf{v}_k|\mathbf{a}_k, \mathbf{z}_i) + \log \frac{p(\mathbf{z}_i)}{q(\mathbf{z}_i|\omega_i)} \right]
\end{equation}

For simplicity, let us assume $p(\mathbf{z}_i) \sim \mathcal{N}(\mathbf{0}, \mathbf{I})$. Considering each task $\omega_i$ is comprised of a support set $\mathcal{S}_i$ and a query set $\mathcal{Q}_i$, the ELBO is reformulated to enhance training efficiency:
 \begin{equation}
\begin{aligned}
    \log &p(\mathbf{v}_{1:|\mathcal{Q}_i|}|\mathbf{a}_{1:|\mathcal{Q}_i|}, \mathcal{S}_i)  \geq \\
    &\mathbb{E}_{q(\mathbf{z}_i|\mathcal{Q}_i)}  \left[  \sum^{\mathcal{Q}_i}_{k=1} \log p(\mathbf{v}_{k}|\mathbf{a}_{k}, \mathbf{z}_i)  +  \text{KL}(q(\mathbf{z}_i|\mathcal{Q}_i) || q(\mathbf{z}_i|\mathcal{S}_i)) \right]
\end{aligned}
\end{equation}
Here, the first term is interpreted as the reconstruction loss for 3D facial animation, and the second term serves as a regularization term, fostering consistency between the support set and the query set.

\subsection{Loss Function}
To train MetaFace, we employ three distinct loss functions: a reconstruction loss, a velocity loss, and a Lnp Loss. 

The reconstruction loss measures the distance between the predicted facial motion bias $V_{\text{pred}}$ and the ground truth $V_{\text{gt}}$, as shown in Equation \ref{equ:recon}.
\begin{equation}
    \label{equ:recon}
    \tau_{\text{recon}} = \frac{1}{N} \sum_{n=1}^{N} (V_{\text{pred}, n} - V_{\text{gt}, n})^2
\end{equation}

To minimize jittery outputs, following \citet{peng2023selftalk}, we use a velocity term to ensure the model learns the facial motion velocity:

\begin{equation}
    \label{equ:vel}
    \tau_{\text{vel}} = \frac{1}{T-1} \sum_{t=2}^T \left(\frac{1}{N} \sum_{n=1}^N (V_{\text{pred}, n}^t - V_{\text{pred}, n}^{t-1} - (V_{\text{gt}, n}^t - V_{\text{gt}, n}^{t-1}))^2\right)
\end{equation}

For the Lnp loss, as previously discussed, the loss function $\tau_{\text{lnp}}$ can be formulated as:
\begin{equation}
    \label{equ:Lnp}
    \tau_{\text{lnp}} = \sum_i p(z_i \mid a_Q) \log\left(\frac{p(z_i \mid a_Q)}{p(z_i \mid a_S)}\right)
\end{equation}

The total loss function can be formulated as:
\begin{equation}
    \label{equ:total_loss}
    \tau = w_1 \tau_{\text{recon}} + w_2 \tau_{\text{vel}} + w_3 \tau_{\text{lnp}}
\end{equation}
where $w_1$, $w_2$, and $w_3$ are hyperparameters that indicate the weight of each corresponding loss component.

\begin{table*}[ht]

    \centering
    \scalebox{0.95}{
    \begin{tabular}{cccccccccc}
    \toprule
        Method & \multicolumn{4}{c}{VOCASet} & &\multicolumn{4}{c}{BIWI} 
        \\ \cline{2-5}\cline{7-10}
        ~ & $L2_{Face}\downarrow$ & $L2_{lip}\downarrow$ & $L2_{max}\downarrow$ & $lip_{sync}\downarrow$ && $L2_{face}\downarrow$ & $L2_{lip}\downarrow$ & $L2_{max}\downarrow$ & $lip_{sync}\downarrow$  \\ 
       \midrule
        VOCA\citep{cudeiro2019capture} & 7.02 & 7.84 & 10.26 & 7.23 && 29.20 & 30.28 & 77.28 & 30.36  \\ 
        FaceFormer\citep{fan2022faceformer} & 1.08 & 5.18 & 9.96 & 4.00 && 11.92 & 12.71 & 35.44 & 12.43  \\ 
        FaceFormer\citep{fan2022faceformer}\dag & 0.85 & 3.24 & 6.62 & 2.95 & &11.56 & 12.33 & 33.94 & 12.05 \\ 
        SelfTalk\citep{peng2023selftalk} & 1.07 & 2.74 & 7.06 & 2.54 && 11.65 & 12.52 & 33.53 & 12.23\\
        SelfTalk\citep{peng2023selftalk}\dag& 0.82 & 2.61 &6.02  & 2.53 & & 10.52 & 11.06 & 31.52& 11.30 \\
        StyleTalk\citep{song2024talkingstyle} & 0.95 & 4.22 & 8.37 & 3.04 && 13.19 & 13.66 & 37.57 & 13.64  \\
        StyleTalk\citep{song2024talkingstyle}\dag& 0.89 & 3.71 & 7.66 & 2.65 & &12.42 & 12.54 & 36.18 & 12.54 \\ 
        Imitator\citep{thambiraja2023imitator}\dag & 0.90 & 2.09 & 5.28 & 1.72 & &- & - & - & -  \\
        Ours\dag & \textbf{0.62} & \textbf{1.86} & \textbf{4.43} & \textbf{1.56} && \textbf{9.15} & \textbf{9.81} & \textbf{26.33} & \textbf{9.49}  \\ \bottomrule
            \end{tabular}}
            \vspace{-0.1in}
    \caption{Qualitive results on VOCASet\citep{cudeiro2019capture} and BIWI-Test-A\citep{BIWI}.
    The sign \dag means the method is retrained on test subjects' train sentences following \citet{thambiraja2023imitator}. The "-" means the method doesn't provide the matched code.  The units of the numbers in the table are all in millimeters (mm).
    }
    \label{tab.main}
    \vspace{-0.2in}

\end{table*}
\section{Experiments}

\subsection{Experimental details}
To assess the efficacy of MetaFace, we conduct experiments using two publicly recognized datasets: VOCASet \citep{cudeiro2019capture} and BIWI \cite{BIWI}. VOCASet includes 480 sentences, each lasting between three to four seconds, captured from 12 subjects using 4D scans at a rate of 60 fps. The BIWI dataset contains 15,000 frames featuring 20 subjects engaged in speech, captured at 25 fps. We sample the audio at 16 kHz and extract audio features using the wav2vec model \citep{baevski2020wav2vec}. In our meta-learning framework, we select an 11-way 1-shot pretraining strategy. The global learning rate is set at $\alpha = 1 \times 10^{-4}$, and the fine-tuning learning rate is $\beta = 5 \times 10^{-5}$. Loss weights are assigned as follows: $w_{\text{recon}} = 1000$, $w_{\text{vel}} = 1000$, and $w_{\text{Lnp}} = 10$. During the task-specific fine-tuning stage, we use four samples from the training sentences of the test subjects, following the methodology described in the imitator model. For evaluation, we measure the L2 vertex error across the entire face ($l2_{\text{face}}$) and within the lip regions ($l2_{\text{lip}}$) to assess the accuracy of the facial animations. Additionally, we employ Dynamic Time Warping (DTW) to evaluate lip synchronization ($lip_{\text{sync}}$) as per the method outlined by \citet{thambiraja2023imitator}. We also gauge performance using the mean of the maximum error per frame ($lip_{\text{max}}$), following the approach described by \citet{richard2021meshtalk}.

\subsection{Quantitative  Evaluation}
% To evaluate the performance of MetaFace, we use L2 vertex error over the entire faces($l2_{face}$) and the lip regions($l2_{lip}$ to evaluate the accuracy of face animation results and use  Dynamic Time Warping(DTW) to evaluate the lip  synchronization($lip_{sync}$) following 
% \citet{thambiraja2023imitator}. Also, we also use mean of the max error per-frame($lip_{max}$) to evaluate the performance following \citet{richard2021meshtalk}. As shown in Tab.\ref{tab.main}, our method achieves both the lower  face animation error and the lip synchronization compared with all other methods. This state-of-the-art results shows the great personality on both whole face motions and lip motions. In particularly, our methods achieves 31\% decrease on face motion distance  and 9.2\% decrease on lip synchronization distances on VOCASet compared with existing personalized talking face generating method \citet{thambiraja2023imitator}. For BIWI dataset, MetaFace achieves 30\% decrease on whole face motion error and 22\% decrease on lip synchronization error.

We compare MetaFace with leading-edge methodologies such as VOCA \citep{cudeiro2019capture}, FaceFormer \citep{fan2022faceformer}, SelfTalk \citep{peng2023selftalk}, StyleTalk \citep{song2024talkingstyle}, and Imitator \citep{thambiraja2023imitator}. For VOCA, it is trained on the BIWI dataset employing the official implementation. FaceFormer, SelfTalk, and StyleTalk undergo external experiments, specifically fine-tuned on four sentences arbitrarily selected from the training set of the test subjects, in alignment with the practices delineated by \citet{thambiraja2023imitator}. Similarly, MetaFace is trained on four samples from the training set of the test subjects. As delineated in Table \ref{tab.main}, MetaFace outperforms all compared methods in terms of reduced facial animation errors and enhanced lip synchronization. These exemplary results signify substantial enhancements in both overall facial movements and lip motions. More precisely, MetaFace registers a 31\% reduction in facial motion distance and a 9.2\% reduction in lip synchronization distances on the VOCASet, surpassing the personalized talking face generation benchmarks set by \citet{thambiraja2023imitator}. On the BIWI dataset, MetaFace achieves a 30\% reduction in whole face motion error and a 22\% improvement in lip synchronization error.

\begin{figure*}[ht]
\centering
\includegraphics[width=0.96\linewidth]{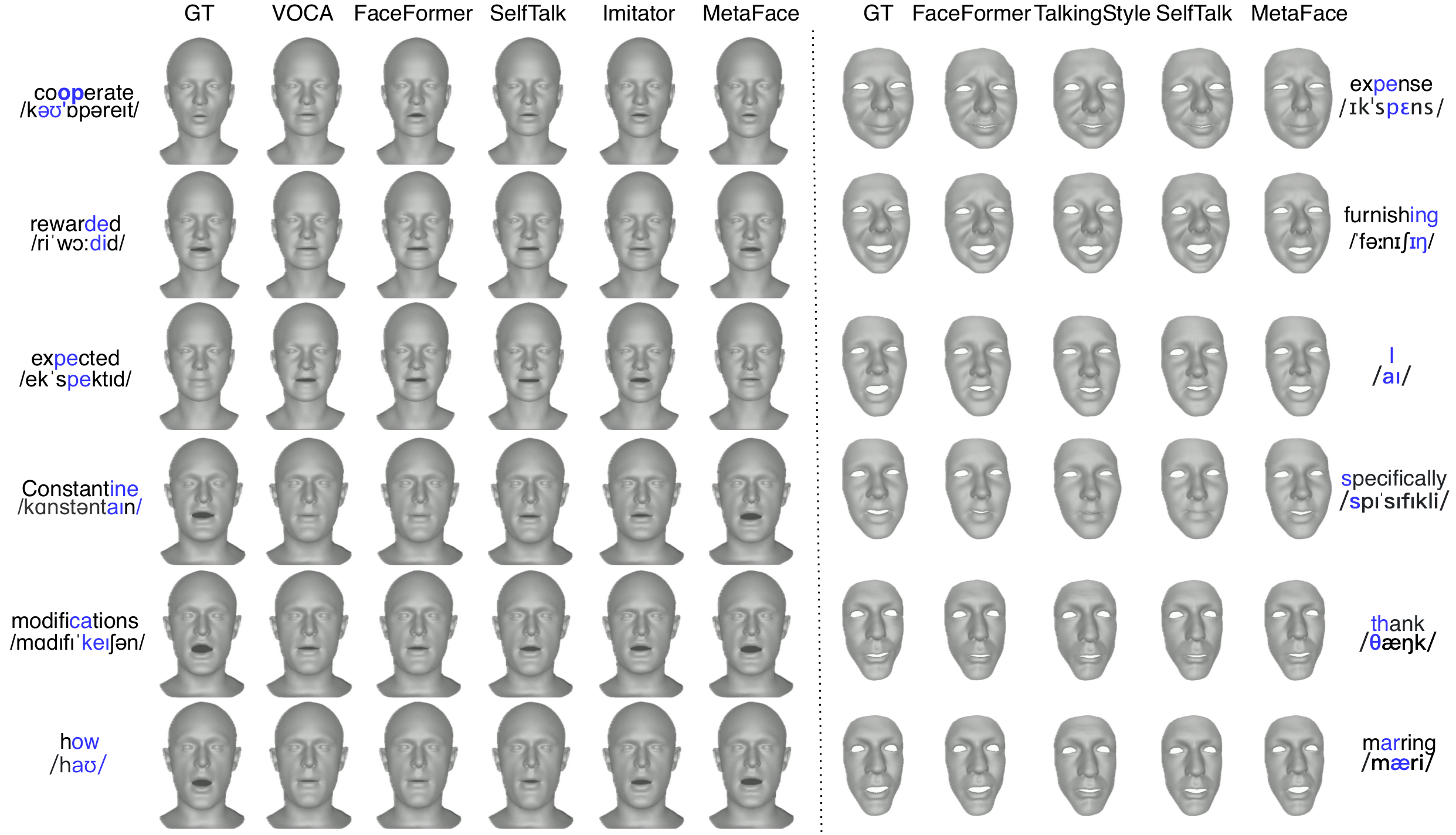}
\caption{Visual comparisons of facial movement by different methods on VOCA-Test (left) and BIWI-Test-A (right).}
\label{fig.example}
\vspace{-0.2in}
\end{figure*}

\subsection{Qualitative Evaluation}
% Although evaluation metrics can effectively demonstrate the realism of a facial model, visual appeal is also a crucial factor in determining the practical application of 3D faces. In the test process, we use the audio in the test set and train the MetaFace model with the first four sentences of test subjects in the training set. We compare VOCA, FaceFormer,  SelfTalk and Imitator on VOCASet and compare with FaceFormer, SelfTalk and TalkingStyle on BIWI dataset. 
% All methods are finetuned with four samples of target person. The images are extracted from the videos in our supplementary material.
% From Fig.\ref{fig.example}, it is easy to figure that the The MetaFace outstanding existing methods with visual results. Specially, with the pronunciation /\textipa{@}u/, MetaFace shows the most likely face motions with mouth shaped like a semicircle and slightly protrudes forward. For closed syllables like /pe/ and /p\textipa{E}/, the mouth shape movements of MetaFace are significantly smaller than those of other methods. For open syllables like /\textipa{a}i/ and /ae/, our method is closer to the ground truth results.

In addition to the quantitative comparisons, we also present a series of visualization results in Fig.\ref{fig.example}. We conduct comparisons of our model, MetaFace, against established methods such as VOCA, FaceFormer, SelfTalk, and Imitator on the VOCASet, and with FaceFormer, SelfTalk, and TalkingStyle on the BIWI dataset. Each method is fine-tuned using four samples from the target individual. The images used for this comparative analysis are extracted from videos included in our supplementary materials. As illustrated in Fig.\ref{fig.example}, MetaFace evidently surpasses the competing methods in terms of visual outcomes. Notably, when articulating the sound /\textipa{@}u/, MetaFace exhibits the most lifelike facial movements, with the mouth forming a semicircle and slightly protruding forward. For closed syllables such as /pe/ and /p\textipa{E}/, the movements of the mouth shapes with MetaFace are markedly more restrained compared to other methods. Similarly, for open syllables like /\textipa{a}i/ and /ae/, our method aligns more closely with the ground truth results

\subsection{User study}

% To investigate the application ability of 3d face animation, we conduct user study on existing dataset following FaceFormer\citep{fan2022faceformer}, Imitator\citep{thambiraja2023imitator} and SelfTalk\citep{peng2023selftalk}. We place the videos side-by-side and require users to select the best one and compute the support radio for each method. We compare MetaFace with SelfTalk and Imitator, which are current best model for personalized talking face animation and lip synchronization. Tab.\ref{tab.user_study} shows that our method achieves greater lip synchronization and realism compared with other models. For instance, our methods achieves 66.4\% support radio compared with Imitator on VOCASet.

To explore the practical efficacy of 3D face animation, we conduct a user study using existing datasets, adhering to the methodologies established by FaceFormer \citep{fan2022faceformer}, Imitator \citep{thambiraja2023imitator}, and SelfTalk \citep{peng2023selftalk}. In this study, participants view videos from each method side-by-side and select the most realistic animation. We compute the support ratio for each method to quantify preference. MetaFace is compared against SelfTalk and Imitator, which are recognized as leading models in personalized talking face animation and lip synchronization. As depicted in Table \ref{tab.user_study}, our method surpasses other models in terms of lip synchronization and overall realism. Notably, MetaFace achieves a support ratio of 66.4\% against Imitator on the VOCASet, emphasizing its superior performance in realistic facial animation.
\begin{table}[!ht]
    \centering
    \scalebox{0.75}{
    \begin{tabular}{ccccc}
\toprule
        Method & \multicolumn{2}{c}{VOCASet}  &\multicolumn{2}{c}{BIWI-Test-B}\\
        ~ & competitor & MetaFace & competitor & MetaFace \\ \midrule
       \textbf{Imitator V.S. MetaFace }& ~ & ~ & ~ & ~ \\ 
        lip sync & 35.3\% & \textbf{64.7\%} & 37.9\% & \textbf{52.1\%} \\ 
        realisim & 38.8\% & \textbf{51.2\%} & 40.4\% & \textbf{59.6\%} \\ \midrule
        \textbf{SelfTalk V.S. MetaFace} & ~ & ~ & ~ & ~ \\ 
        lip sync & 33.6\% & \textbf{66.4\%} & 35.1\% & \textbf{64.9\%} \\ 
        realisim & 37.2\% & \textbf{62.8\%} & 36.2\% & \textbf{63.8\%} \\ \bottomrule
    \end{tabular}}
     \vspace{-0.1in}
    \caption{User study on VOCASet and BIWI-Test-A.}
    \label{tab.user_study}
     \vspace{-0.1in}
\end{table}

\subsection{Ablation Study}

In this section, we conduct an ablation study to demonstrate the effectiveness of our key design components. The experiments are conducted on the VOCASet, with results presented in Table \ref{tab:ablation}.

\noindent \textbf{Impact of the Robust Meta Initialization Stage:} The primary objective of MetaFace is to learn a meta face that serves as an initialization for all faces. As illustrated in the third row of Table \ref{tab:ablation}, the robust meta initialization contributes significantly to the overall performance. Specifically, this meta-learning approach results in a 24.4\% decrease in $l2_{face}$ and a 25.7\% decrease in $lip_{sync}$.

\noindent \textbf{Impact of the Dynamic Relation Mining Neural Process:} This component enables MetaFace to discern the relationship between observed speaking styles and unobserved ones. As shown in Table \ref{tab:ablation}, the contribution of the Neural Process primarily leads to a 14.4\% reduction in the maximum $l2$ distance within the lip region and an 11\% reduction in lip synchronization distance.

\noindent \textbf{Impact of the Low-rank Matrix Memory Reduction Approach} As indicated in the first row of Table \ref{tab:ablation}, the model excluding LoRA (Low-Rank Adaptation) achieves the most favorable outcomes. While utilizing the full model parameters delivers optimal performance, it concurrently necessitates a hundredfold increase in model parameter training requirements. The integration of the LoRA model substantially reduces trainable parameters by approximately 99.5\%, with a modest performance decrease not exceeding 7\%. From the second row of Table \ref{tab:ablation}, it is evident that meta-learning plays a crucial role in enhancing outcomes. Meta training of the low-rank decomposition module improves lip synchronization by 17\% and reduces face motion error by 16\%.

\begin{table}[htbp]
    \centering
    \scalebox{0.9}{
    \begin{tabular}{lccccc}
    \toprule
        Method & $l2_{face}$ & $l2_{lip}$ & $l2_{max}$ & $l2_{sync}$ & Params \\
    \midrule
        w/o LoRA & \textbf{0.61} & \textbf{1.77} & \textbf{4.21} & \textbf{1.47} & 340.4MB \\ 
        w/o LoRA Meta & 0.73 & 2.06 & 4.69 & 1.88 & 1.94MB \\ 
        w/o RMIS & 0.83 & 2.12 & 4.76 & 1.85 & 1.94MB \\ 
        w/o DRMN & 0.76 & 2.09 & 5.18 & 1.75 & 1.94MB \\ 
    \midrule
        MLFaces & \underline{0.62} & \underline{1.86} & \underline{4.43} & \underline{1.56} & \textbf{1.94MB} \\ 
    \bottomrule
    \end{tabular}
    }
     \vspace{-0.1in}
    \caption{Ablation study of MLFaces on VOCASet. The optimal values are \textbf{bolded}, while the second-best values are \underline{underlined}.}
    \label{tab:ablation}
    \vspace{-0.1in}
\end{table}

\begin{table}[ht]
    \centering
    \begin{tabular}{ccccc}
    \toprule
         \textbf{Number} & $l2_{face}$ & $l2_{lip}$ & $l2_{max}$ & $l2_{sync}$ \\ \midrule
        1 & 0.96 & 2.07 & 5.64 & 1.81 \\ 
        2 & 0.88 & 1.97 & 5.24 & 1.74 \\ 
        3 & 0.73 & 1.89 & 4.92 & 1.67 \\ 
        4 & \textbf{0.62} & \textbf{1.86} & \textbf{4.43} &\textbf{1.56} \\ \bottomrule
    \end{tabular}
    \caption{Results of different samples for specific person speaking style.  }
    \label{tab.number}
    \vspace{-0.1in}
\end{table}

\subsection{Generalization Test towards Adaptation Samples}
We also investigate the generalization ability of our model concerning the number of adaptation samples. In our experiments, we vary the number of seen sample sequences and train the model with randomly selected sentences. Results, shown in Table \ref{tab.number}, indicate that increasing the sample count significantly reduces the $l2_{face}$ error. Specifically, using four sentences decreases the error by up to 36\%, demonstrating that a greater number of samples notably improves the model’s generalization capability in facial feature reconstruction.

\section{Conclusion}
In this paper, we have introduced MetaFace, a novel model for audio-driven 3D face animation that  addresses the challenge of personalized speaking style adaptation. MetaFace comprises three key components: the Robust Meta Initialization Stage (RMIS), which facilitates fundamental speaking style adaptation; the Dynamic Relation Mining Neural Process (DRMN), which establishes connections between observed and unobserved speaking styles; and the Low-rank Matrix Memory Reduction Approach, which enhances the efficiency of model optimization and the learning of style details. By integrating these novel designs, MetaFace not only significantly surpasses robust existing baselines but also sets a new benchmark in the field.

\bibliography{main}

\end{document}